\documentclass[
]{ceurart}

\usepackage{listings}
\begin{document}

\copyrightyear{2024}
\copyrightclause{Copyright for this paper by its authors.
  Use permitted under Creative Commons License Attribution 4.0
  International (CC BY 4.0).}

\conference{CHIRP 2024: Transforming HCI Research in the Philippines Workshop, May 09, 2024, Binan, Laguna}

\title{Alexa, I Wanna See You: Envisioning Smart Home Assistants for the Deaf and Hard-of-Hearing}



\author[]{Tyrone Justin {Sta. Maria}}[%
email=tyrone_stamaria@dlsu.edu.ph,%
]
\author[]{Jordan Aiko {Deja}}[%
email=jordan.deja@dlsu.edu.ph,%
]

\address[1]{De La Salle University, Manila, Philippines}

\begin{abstract}
Smart Home Assistants (SHAs) have become ubiquitous in modern households, offering convenience and efficiency through its voice interface. However, for Deaf and Hard-of-Hearing (DHH) individuals, the reliance on auditory and textual feedback through a screen poses significant challenges. Existing solutions primarily focus on sign language input but overlook the need for seamless interaction and feedback modalities. This paper envisions SHAs designed specifically for DHH users, focusing on accessibility and inclusion. We discuss integrating augmented reality (AR) for visual feedback, support for multimodal input, including sign language and gestural commands, and context awareness through sound detection. Our vision highlights the importance of considering the diverse communication needs of the DHH community in developing SHA to ensure equitable access to smart home technology.
\end{abstract}


\maketitle

\section{Introduction}

\begin{figure}
    \centering
    \includegraphics[width=1\linewidth]{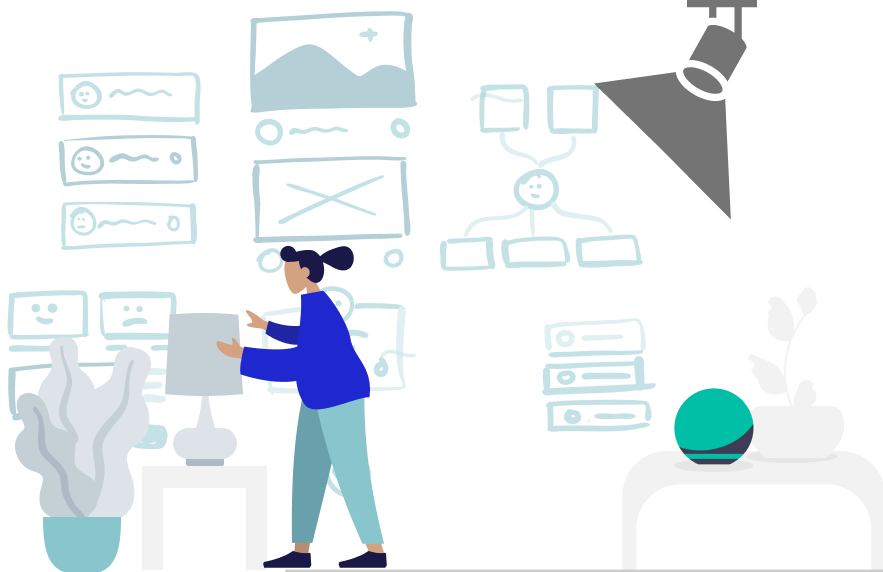}
    \caption{The vision of a smart home assistant designed for deaf and hard-of-hearing individuals that utilizes augmented reality for visual output, has multiple modes of input, and has context awareness.}
    \label{fig:pos-paper-fig}
\end{figure}


\par Smart Home Assistants (SHAs), such as Amazon Alexa and Google Home, have become increasingly integrated into our daily routines, offering seamless augmentation of various activities. For instance, users can ask the SHA a question rather than manually searching for information on a device and receive an immediate, voice-based response. This interaction, typically conducted through voice commands, allows users to focus on their tasks without diverting their attention to a screen.

\par However, this convenience is not equally accessible to Deaf and Hard-of-Hearing (DHH) individuals, who could otherwise benefit from the efficiencies offered by SHAs. The primary mode of interaction—voice commands—poses significant challenges for DHH users. Due to their inability to hear their speech, DHH individuals often experience difficulties with speech production, leading to reduced speech intelligibility and quality~\cite{thirumalai2004speech}. This challenge is compounded by the fact that speech recognition software is typically trained on the speech patterns of hearing individuals~\cite{glasser2017feasibility}. As a result, DHH users encounter substantial barriers when attempting to interact with SHAs. For example, research by Glasser et al.~\cite{glasser2017feasibility, glasser2017deaf} found that DHH users experienced a 77\% word error rate when using automatic speech recognition software, compared to an 18\% error rate for hearing users. These disparities highlight the need for more inclusive design approaches that enable DHH individuals to fully engage with SHAs.

\par SHAs predominantly provide feedback through spoken responses and visual indicators, such as lights, which create significant accessibility challenges for DHH users~\cite{blair2020didn}. Some devices, like the Amazon Echo Show\footnote{https://www.amazon.com/echo-show-10/dp/B07VHZ41L8}, attempt to address these issues by incorporating a screen, allowing DHH users to interact with SHAs via touch gestures and consume information visually. However, this solution disrupts the seamless interaction typically facilitated by SHAs' voice-based conversational outputs, as users must shift their attention to read the feedback on the screen~\cite{blair2020didn}.

\par To address these problems, researchers have opted to explore sign language as an input modality for SHAs~\cite{glasser2021understanding, glasser2022analyzing}. DHH individuals interacted with SHAs that understand sign language using sign language and finger spelling~\cite{glasser2022analyzing}. This approach is typical when communicating proper nouns or words without corresponding signs, reflecting a common practice within the community. Common commands used by these individuals involved looking up information, setting alarms, connecting with other smart devices, and adding captions to videos~\cite{glasser2021understanding, glasser2022analyzing}. 

\par For wake-words (e.g. \textit{Alexa}, \textit{Hey Google}), they typically waved in front of the device as it is the common form of getting the attention of others when they want to communicate~\cite{mande2021deaf, glasser2022analyzing} or they finger-spelled the exact word. For showing feedback, DHH individuals prefer to get results through text output from a screen~\cite{glasser2021understanding}. 

\par Existing solutions primarily focused on providing input to SHAs using sign language. Current feedback modalities that can be used comfortably by DHH individuals are text output from a screen. This interrupts the seamless interaction facilitated by SHAs' conversational outputs, as these individuals must divert their attention in reading the feedback from the screen~\cite{blair2020didn}. Thus, in this paper, we envision ways in which DHH users can interact seamlessly with SHAs. We argue that such design approaches may provide a direction for considering the needs of DHH users in future iterations of SHAs.

\section{Smart Home Assistant for DHH}

\par We present three approaches to make better Smart Home Assistants for DHH. Namely, (i) using AR, (ii) integrating multimodal input and (iii) having context-aware spaces. 

\subsection{Using Augmented Reality} \label{sec:using-ar}

\par While seeing users in their homes typically navigate their environment easily, designing Smart Home Assistants (SHAs) for Deaf and Hard-of-Hearing (DHH) individuals requires special consideration to ensure they can multitask effectively. The goal is to enable these users to receive and process information without diverting their attention to a specific device. One approach is incorporating augmented reality (AR), which provides visual feedback directly within the user's physical space. This can be achieved by integrating mobile steerable projectors, as described in~\cite{cauchard2011mobile, lightform, mathew2024deaf}. Using projectors eliminates the need for additional wearable accessories, which can be uncomfortable for DHH users, particularly those who wear hearing aids~\cite{mathew2024deaf}. By embedding AR into SHAs, DHH individuals can benefit from visual feedback that remains within their field of view, allowing them to stay engaged with their surroundings without being tethered to the SHA device. For instance, ingredient lists could be projected nearby when cooking, or a timer notification could be displayed even if the user has stepped away from the stove. Such enhancements would significantly improve their ability to consume information and complete tasks seamlessly and efficiently.

\subsection{Multimodal Input}
\par In designing smart home assistants (SHAs) for individuals who are deaf and hard-of-hearing (DHH), it is essential to consider their preferred mode of communication. Supporting multiple input modes can help cater to their needs, whether sign language~\cite{mitchell2006many} or verbal communication with the help of hearing aids~\cite{blair2020didn, glasser2017deaf}. When interacting with SHAs, users typically begin with invoking a wake-word followed by a query, with responses delivered through auditory feedback, on-screen text, or device control. However, when using sign language, certain words or expressions lack direct translation, leading to finger spelling, which takes more effort than verbal communication. For example, DHH individuals used finger spelling to express the wake-word ``Alexa''~\cite{glasser2022analyzing}. Using gestural inputs should be explored to lessen their efforts on such tasks~\cite{mande2021deaf}. We could extend this to menial tasks such as asking for the weather, setting alarms, and controlling other smart devices. For example, users can clap once to turn on a smart light or show their palm to ask for the weather. By considering a range of input modes, SHAs can be designed to be more accessible and user-friendly for DHH individuals.

\subsection{Context Aware Spaces}

\par Deaf and Hard-of-Hearing (DHH) individuals often rely on voice-to-text programs to aid in understanding speech, extending their auditory perception~\cite{glasser2017deaf, glasser2017feasibility}. This concept can be further expanded to Smart Home Assistants (SHAs) by enabling them to monitor and interpret household sounds, such as the whistle of boiling water, door knocks, or the activity of appliances~\cite{laput2018ubicoustics, wu2020automated}. By understanding and responding to these auditory cues, SHAs can offer more context-aware assistance, extending their functionality beyond traditional tasks like internet searches, setting alarms, or providing weather updates. Additionally, integrating augmented reality (AR) to deliver visual alerts and information can significantly enrich the daily experiences of DHH individuals, especially in situations that depend on audio cues. For example, an SHA could notify a DHH user of a boiling kettle while they are occupied with another task or alert them to someone knocking at the door, thereby enhancing their awareness and interaction with their surroundings.

\section{Discussion}
\par We based the visions of a SHA designed for DHH individuals to make the technology accessible and augment their daily activities. SHAs have become ubiquitous and are becoming integrated into our daily lives~\cite{edwards2001home}. We introduce ways for DHH individuals to interact with SHAs seamlessly, from simply an information provider and smart device controller to an extension of their hearing around their living space. We believe that the features we presented may represent that of a SHA designed with DHH users in mind, however this is subject to further exploration and validation with actual users. We also discuss some benefits and disadvantages that go with it as well. 

\par SHAs usually have smart speakers and a mic that recognizes voice inputs and outputs audio feedback. However, these modalities are not accessible to DHH individuals~\cite{thirumalai2004speech, glasser2017feasibility}. This is why the SHA designed for DHH users must attempt to become a companion that can extend their senses to facilitate seamless interaction with smart home technology alongside their overall home environment.

\par It will also be equally challenging to set up an AR environment within the home~\cite{lamberti2014challenges}. SHAs usually come out of the box with smart speakers that are sometimes accompanied by a display without any additional setup. Using a steerable projector (e.g. such as in the work of ~\cite{cauchard2011mobile}) comes up with challenges in terms of its placement~\cite{park2019deep}. It would be ideal for the projection to capture every corner of the room without being obstructed. However, achieving the ideal placement requires additional equipment and expertise~\cite{speicher2018xd}. 
 
\par In the design of SHAs for DHH, we factored in several modalities and parameters, such as image data for sign language detection and sound data detecting voice and the home environment~\cite{tran2024assessment, blair2020didn}. Security and privacy must be strictly upheld in all these conditions, especially if it involves data on users' activity~\cite{pradhan2018accessibility}. As we factor in more considerations in designing SHAs for DHH in the future, we introduce more benefits and issues with them. We posit that SHAs designed for DHH must always consider the unique communication needs and preferences of the DHH community ~\cite{blair2019understanding} while prioritizing accessibility, inclusivity~\cite{jain2019exploring}, and robust security measures to safeguard users' sensitive data.
 
\section{Summary}
\par We outlined the challenges that DHH individuals face when interacting with SHAs. We looked at existing solutions that address these problems. Through our examination of the existing challenges and current solutions, we have formulated a vision for SHAs designed for DHH users. Our vision prioritizes accessibility. Incorporating features such as AR integration, support for multimodal input, and context awareness. Furthermore, we emphasize the importance of considering the unique communication needs and preferences of the DHH community for both input and output modalities when developing SHAs aligned with DHH users.

\bibliography{main}




\end{document}